# Probing Carrier Dynamics in *sp*$^3$-Functionalized Single-Walled Carbon Nanotubes with Time-Resolved Terahertz Spectroscopy

*Wenhao Zheng*[1#]*, Nicolas F. Zorn*[2#]*, Mischa Bonn*[1]*, Jana Zaumseil*[2*]*, Hai I. Wang*[1*]

[1] Max Planck Institute for Polymer Research, D-55128 Mainz, Germany

[2] Institute for Physical Chemistry and Centre for Advanced Materials, Universität Heidelberg, D-69120 Heidelberg, Germany

Corresponding Authors

*E-mail: wanghai@mpip-mainz.mpg.de

*E-mail: zaumseil@uni-heidelberg.de

ABSTRACT

The controlled introduction of covalent $sp^3$ defects into semiconducting single-walled carbon nanotubes (SWCNTs) gives rise to exciton localization and red-shifted near-infrared luminescence. The single-photon emission characteristics of these functionalized SWCNTs make them interesting candidates for electrically driven quantum light sources. However, the impact of $sp^3$ defects on the carrier dynamics and charge transport in carbon nanotubes remains an open question. Here, we use ultrafast, time-resolved optical-pump terahertz-probe spectroscopy as a direct and quantitative technique to investigate the microscopic and temperature-dependent charge transport properties of pristine and functionalized (6,5) SWCNTs in dispersions and thin films. We find that $sp^3$ functionalization increases charge carrier scattering, thus reducing the intra-nanotube carrier mobility. In combination with electrical measurements of SWCNT network field-effect transistors, these data enable us to distinguish between contributions of intra-nanotube band transport, $sp^3$ defect scattering and inter-nanotube carrier hopping to the overall charge transport properties of nanotube networks.

KEYWORDS

(6,5) single-walled carbon nanotubes, luminescent $sp^3$ defects, terahertz spectroscopy, photoconductivity, field-effect transistors, charge scattering

Highly purified semiconducting single-walled carbon nanotubes (SWCNTs) are a promising material for electronic circuits and optoelectronic devices owing to their high charge carrier mobilities and narrowband near-infrared (nIR) emission.[1-3] Recently, the controlled functionalization of SWCNTs with covalently attached aryl or alkyl groups has emerged as a versatile approach to enhance their optical properties.[4-7] These local $sp^3$-hybridized lattice defects, also referred to as quantum defects or organic color centers, form new electronic states that efficiently trap excitons, leading to red-shifted (by ~100–300 meV) nIR photoluminescence (PL) with increased quantum yields.[8-12] Furthermore, such functionalization enables high-purity single-photon emission in SWCNTs at room temperature.[13-14] These $sp^3$ defects also influence the intrinsic charge transport along the nanotubes (*i.e.*, intra-nanotube band transport[15]) as indicated by the reduced conductivities that were found for individual functionalized SWCNTs (both metallic and semiconducting).[16-17] In contrast to single nanotubes, charge transport in SWCNT networks is commonly believed to be limited by thermally activated carrier hopping across the nanotube-nanotube junctions,[18] which leads to their substantially lower field-effect mobilities of only 5–50 cm$^2$ V$^{-1}$ s$^{-1}$ compared to around 1000 cm$^2$ V$^{-1}$ s$^{-1}$ for a single nanotube depending on its diameter.[19] Nevertheless, recent studies have shown that the introduction of $sp^3$ defects also lowers the overall charge carrier mobility in dense nanotube networks.[20] The apparent convolution of intra- and inter-nanotube transport in a network (see **Figure 1a**) further complicates the interpretation of the impact of $sp^3$ defects. Consequently, experimental techniques are required that can determine local charge transport and carrier mobilities within individual SWCNTs or SWCNT segments. For example, flash-photolysis time-resolved microwave conductivity on nanotube dispersions[21] and dark microwave conductivity measurements on chemically doped nanotube networks[22] have provided some initial insights for pristine SWCNTs but were not applied to functionalized nanotubes.

Ultrafast optical-pump terahertz (THz)-probe (OPTP) spectroscopy is another contact-free, all-optical method that has been shown to enable a quantitative description of intrinsic charge transport properties (*e.g.*, charge scattering time) of various nanomaterials in dispersions and films, including pristine carbon nanotubes.[23-27] Due to the transient nature of the THz probe with an oscillating electrical field with ~1 ps duration, charge carriers are locally driven over a distance of tens of nm, rendering THz spectroscopy a highly suitable technique to probe charge transport on a microscopic level.[23, 28] Previous THz studies revealed significantly longer charge scattering times for (6,5) SWCNTs compared to graphene nanoribbons,[24] as well as very similar carrier dynamics for semiconducting and metallic SWCNT films.[29]

Here, we investigate the intrinsic charge transport properties of $sp^3$-functionalized, polymer-sorted (6,5) SWCNTs as a model system by combining (temperature-dependent) OPTP spectroscopy of dispersions and thin films with electrical measurements of nanotube network field-effect transistors. The observed decrease in magnitude and lifetime of the transient photoconductivity with the degree of functionalization corroborates the impact of $sp^3$ defects on intra-nanotube carrier mobilities and provides a clearer picture of the contributions of intra-nanotube and inter-nanotube charge transport in pristine and functionalized SWCNT networks.

RESULTS AND DISCUSSION

Monochiral, semiconducting (6,5) SWCNTs (diameter ~0.76 nm, average length ~1.4 µm) were obtained by highly selective dispersion of CoMoCAT nanotubes with a fluorene-bipyridine copolymer (PFO-BPy) in toluene.[30] These polymer-wrapped (6,5) SWCNTs were covalently functionalized with 4-nitrobenzenediazonium tetrafluoroborate[31] to create different densities of $sp^3$ defects (see **Figure 1a**) on the nanotubes (for further details, see **Methods**). The degree of functionalization was quantified with various spectroscopic metrics. The introduction of $sp^3$ defects gives rise to a small $E_{11}^*$ defect absorption band at ~1160 nm (see **Figure 1b** and **Supporting Information, Figure S1**) and a significant $E_{11}^*$ PL emission feature at ~1172 nm (**Figure 1c**). Both increase relative to the native $E_{11}$ excitonic transition with increasing defect concentration. Furthermore, resonant Raman spectroscopy (**Figure 1d**) showed an increasing but still relatively low signal of the defect-related D mode in relation to the $G^+$ mode, which is associated with the $sp^2$ carbon lattice.[31] As shown in **Figure S2** (**Supporting Information**), a linear correlation between the Raman $D/G^+$ area ratio and the integrated $E_{11}^*/E_{11}$ area ratios in absorption and emission was found. While all three quantities may serve as metrics for the defect concentration of $sp^3$-functionalized carbon nanotubes,[20, 31] here we will use only the Raman $D/G^+$ ratio to reflect the degree of functionalization in our samples. The estimated $sp^3$ defect densities (based on the differential Raman $D/G^+$ area ratio)[32] range between 5 and 20 defects per µm, corresponding to an average distance of about 200 to 50 nm between two adjacent defects. Note that samples with different defect densities still possess the same length distribution, as determined by atomic force microscopy (see **Supporting Information, Figure S3**).

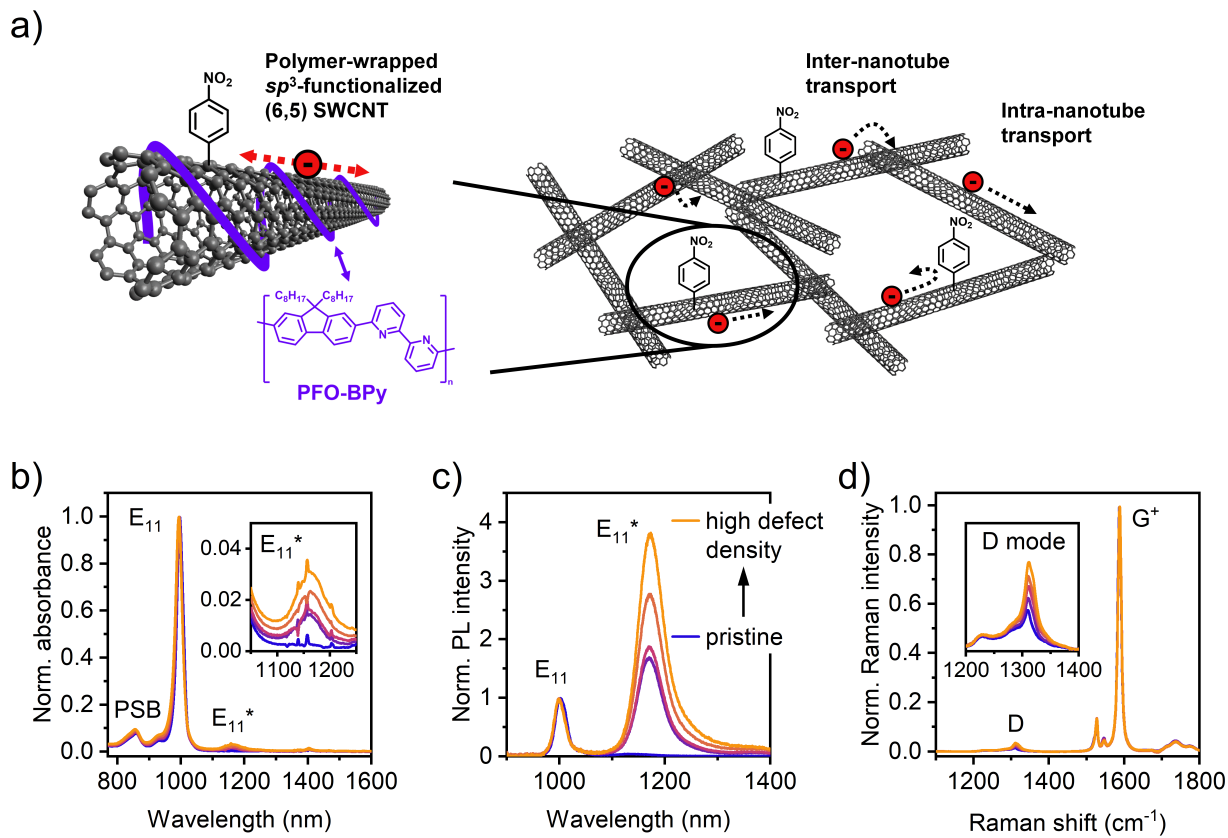

**Figure 1. a)** Schematic of a polymer-wrapped, $sp^3$-functionalized (6,5) SWCNT with the molecular structure of the wrapping polymer PFO-BPy. Charge transport through a network of such SWCNTs involves carrier hopping across nanotube-nanotube junctions (inter-nanotube transport) as well as intra-nanotube transport along the individual SWCNTs. **b)** Normalized absorption spectra of pristine and $sp^3$-functionalized (6,5) SWCNT dispersions. Labels indicate the main excitonic $E_{11}$ transition with its phonon sideband (PSB) and the $E_{11}^*$ defect-state absorption feature. Inset: zoom-in on the $E_{11}^*$ absorption band. **c)** Normalized PL spectra of pristine and $sp^3$-functionalized (6,5) SWCNT dispersions after $E_{22}$ excitation at 575 nm. **d)** Raman spectra of pristine and $sp^3$-functionalized (6,5) SWCNT films normalized to the $G^+$ mode. Inset: zoom-in on the defect-related D mode, which increases with $sp^3$ defect density.

To investigate the impact of *sp*³ functionalization on the photoconductivity dynamics and transport properties of individual polymer-wrapped SWCNTs in dispersion, we employed ultrafast optical-pump terahertz (THz)-probe (OPTP) spectroscopy as schematically shown in **Figure 2a**. In OPTP measurements, inter-band photoexcitation (~1.25 eV optical pump pulses, close to the $E_{11}$ transition peak at ~1000 nm) generates a small population of free charge carriers that are subsequently probed by a single-cycle THz pulse with ~2 THz bandwidth. The field of the incident THz pulse accelerates the free charge carriers, whereby some of the energy in the THz field is dissipated. The degree of dissipation is a direct measure of charge carrier mobility. The transmitted THz field $E$ is characterized directly in the time domain, so that both its amplitude and phase can be determined, thus providing direct access to the complex-valued photoconductivity $\Delta\sigma$. By monitoring the photoinduced THz absorption ($\Delta E = E_{pump} - E$) at different pump-probe delays, the time-dependent photoconductivity ($\Delta\sigma \propto -\frac{\Delta E}{E}$) dynamics can be traced (see the **Supporting Information** for a detailed description).[23]

**Figure 2b** compares the magnitude of the real component of the photoconductivity of three representative nanotube dispersions (pristine, low and high degree of *sp*³ functionalization; for complete datasets, see **Supporting Information, Figure S4**). In all samples, the optical generation of free carriers is manifested by a steep rise in photoconductivity. We note that the generation of free carriers following excitation to the lowest excitonic states was reported previously for polymer-wrapped SWCNTs,[21] yet the detailed mechanism remains elusive. One of the plausible explanations involves the thermodynamic equilibrium between free carriers and excitons (*e.g.*, following Saha equation).[33-35] Further investigation is required for a better understanding of free carrier generation under these conditions, which is beyond the focus of the current study.

As shown in **Figure 2b**, $sp^3$ functionalization results in a pronounced decrease in the peak photoconductivity amplitude for a given absorbed photon density $N_{abs}$ and thus approximately the same photogenerated charge carrier density as for the pristine nanotubes (the photon-to-free-charge conversion quantum yield is expected to be similar in all samples). The nanotube dispersion with the highest $sp^3$ defect concentration exhibits a reduction of ~30 % in $\Delta\sigma$ compared to pristine SWCNTs. In dispersion, OPTP probes individual nanotubes, and thus inter-nanotube electronic coupling can be excluded. This assumption is supported by the nearly identical photoconductivity (both amplitude and dynamics after normalizing to the absorbed photon density) of dispersions with different nanotube concentrations (see **Supporting Information, Figure S5**). As such, our observation corroborates the direct impact of $sp^3$ functionalization on intra-nanotube charge transport. We assign the reduction of photoconductivity to charge scattering at the $sp^3$ defects. The scattering probability increases with increasing defect density, thus accounting for the decreased photoconductivity. For the highest defect density investigated here (~20 defects per µm), the majority of photogenerated carriers interacts with a defect site during their lifetime (see estimation of carrier diffusion length of about 30-50 nm, **Supporting Information, Figure S6**).

After reaching a maximum, the photoconductivity decays rapidly, followed by a slow decay component (see **Figure 2b**). The fast decay can be directly related to the formation of bound electron-hole pairs (*i.e.*, excitons),[24, 33] which do not contribute to the conductance. Exciton localization may occur at $sp^3$ defects. Recent transient absorption studies reported exciton trapping at defects on sub-10 ps time scales for very high defect densities.[36] Here, the impact of $sp^3$ defects on the photoconductivity dynamics, in particular on carrier localization, is shown in the normalized transients in **Figure 2c** and **Supporting Information, Figure S7**. Increasing the $sp^3$ defect density leads to a significant reduction in the photoconductivity lifetimes. The exciton localization probability is enhanced as manifested by the increased weight of the fast decay in the photoconductivity with increasing defect density.

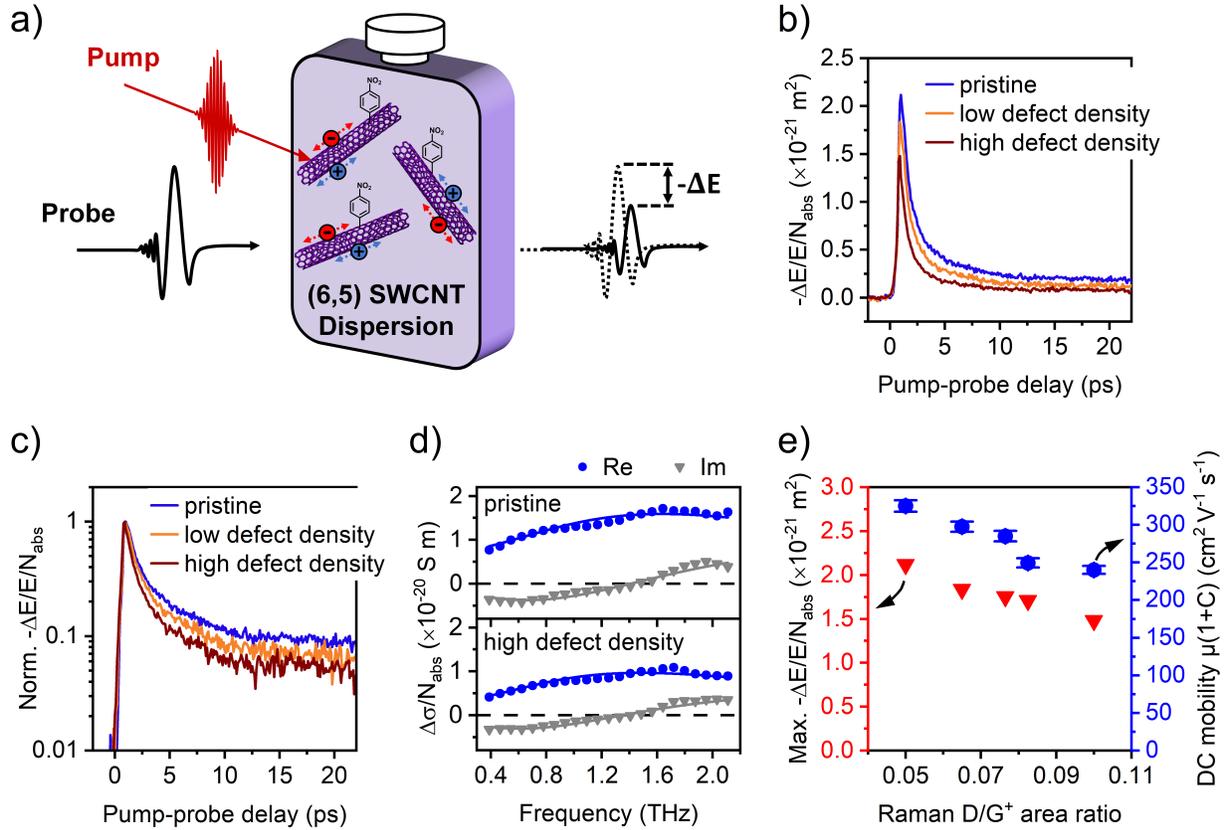

**Figure 2**. **a)** Schematic illustration of optical-pump THz-probe (OPTP) spectroscopy on nanotube dispersions. **b)** Photoconductivity dynamics of pristine and *sp*$^3$-functionalized (low and high defect density) SWCNT dispersions in toluene following optical excitation at 1.25 eV at room temperature (absorbed photon density, ~3·10$^{18}$ m$^{-2}$) and **c)** data normalized to the maximum photoconductivity (logarithmic scale). **d)** Complex photoconductivity spectra of pristine and *sp*$^3$-functionalized (high defect density) SWCNT dispersions measured ~1.5 ps after photoexcitation (blue circles, real conductivity; gray triangles, imaginary conductivity). The solid lines represent fits to the data with the Drude-Smith model. **e)** Maximum photoconductivity (red triangles) extracted from a) and photogenerated carrier mobility in the DC limit (blue circles) determined from fit in d) depending on the *sp*$^3$ defect density (Raman D/G$^+$ area ratio).

To further clarify the impact of $sp^3$ functionalization on the THz photoconductivity in SWCNTs and the nature of charge carriers following optical excitations, we conducted THz time-domain spectroscopy (THz-TDS; for experimental details and data analysis, see **Methods** and **Supporting Information**), from which the frequency-resolved photoconductivity can be obtained. **Figure 2d** illustrates the complex photoconductivity spectra measured at a delay of ~1.5 ps after photoexcitation (see **Supporting Information, Figure S8** for spectra at different delays up to 6 ps). The large real conductivity (along with a small imaginary conductivity) indicates the dominant role of free carriers in the photoconductivity response at early times after optical excitation. We find that the Drude-Smith (DS) model, which was previously applied to describe charge transport in carbon nanotubes,[24, 29] provides a good description of the data. This model describes the transport of free carriers in a medium where charges experience preferential backscattering, *e.g.*, due to nanoscale confinement. The DS model equation reads[37-38]

$$\sigma(\omega) = \frac{\varepsilon_0 \omega_p^2 \tau}{1 - i\omega\tau}\left(1 + \frac{C}{1 - i\omega\tau}\right), \text{ with } \omega_p^2 = \frac{e^2 N}{\varepsilon_0 m^*} \quad (1)$$

where $\tau$, $\omega_P$, $\varepsilon_0$, $e$, and $m^*$ are the effective carrier momentum scattering time, plasma frequency, vacuum permittivity, elementary charge, and charge effective mass, respectively. The parameter $C$ characterizes the probability of backscattering at, *e.g.*, defects or/and grain boundaries. Values for $C$ can range from 0 (isotropic scattering) to -1 (complete backscattering). From the DS parameters, we infer the mobility $\mu = \frac{e \cdot \tau}{m^*}(1 + C)$ in the DC limit, which is shown as a function of the Raman D/G$^+$ area ratio (*i.e.*, $sp^3$ defect density) in **Figure 2e** together with the extracted photoconductivity amplitudes. Both exhibit a very similar trend, and the ~25 % reduction of carrier mobility for the sample with the highest defect concentration is quantitatively consistent with the decrease in photoconductivity amplitude. These values suggest that the impact of *sp*$^3$ defects on intra-nanotube carrier mobility, rather than on photogenerated carrier density, governs the reduction of photoconductivity upon *sp*$^3$ functionalization (see **Figure 2b**). Similarly, we observe an increase in the charge carrier scattering rate of ~20 % induced by the *sp*$^3$ defects for the highest Raman D/G$^+$ area ratio (see **Supporting Information, Figure S9**). These values correlate well with a conductivity decrease of ~20 % associated with the introduction of single defects in individual nanotubes.[16-17] Although the extracted mobilities of several hundred cm$^2$V$^{-1}$s$^{-1}$ are in good agreement with the reported diameter-dependent mobilities for single-nanotube transistors,[19] it should be noted that the carrier mobilities extracted here represent local values (on the length scales of tens of nm) and are measured at low carrier densities (approximately 1 charge carrier per µm of SWCNT). In addition, the extracted mobility values may depend on the applied fitting model (*e.g.*, the Drude-Smith model here). However, the relative trend of carrier mobilities is robust and does not depend on the fitting.

To compare the impact of *sp*$^3$ defects on the local carrier mobilities in individual nanotubes to macroscopic charge transport in networks, we also fabricated field-effect transistors (FETs) with dense, spin-coated films (see **Supporting Information, Figure S10**) of pristine and *sp*$^3$-functionalized (6,5) SWCNTs and interdigitated source-drain electrodes with channel length and width of 20 µm and 10 mm, respectively. **Figure 3a** shows a schematic cross-section of an FET with a representative AFM image of a functionalized nanotube network. Due to the required amount of material, different series of SWCNT dispersions were prepared for OPTP spectroscopy and FET fabrication. However, as indicated by the very similar $E_{11}$*/$E_{11}$ absorption area ratios and Raman D/G$^+$ area ratios (see **Supporting Information, Figure S11**), the degree of *sp*$^3$ functionalization was similar in both cases and thus should not affect the comparison. The full spectroscopic characterization of the SWCNT dispersions used for FET fabrication is shown in **Figure S12** (**Supporting Information**).

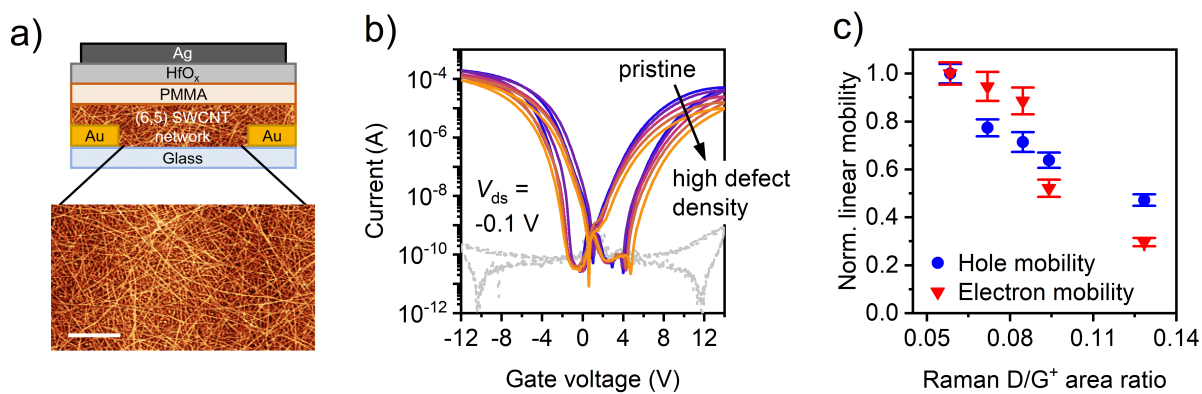

**Figure 3. a)** Schematic illustration of a bottom-contact, top-gate SWCNT network FET (layer thicknesses not to scale) together with a representative atomic force micrograph of a $sp^3$-functionalized (6,5) SWCNT network (scale bar, 500 nm). **b)** Ambipolar transfer characteristics (source-drain voltage $V_{ds}$ = -0.1 V) of FETs based on networks of pristine and $sp^3$-functionalized (6,5) SWCNTs with different defect densities (solid lines, drain currents $I_d$; gray dotted lines, gate leakage currents $I_g$). **c)** Linear charge carrier mobilities (blue circles, hole mobilities; red triangles, electron mobilities) of SWCNT network FETs normalized to the pristine network reference transistors *versus* Raman $D/G^+$ area ratio. Error bars are standard deviations for at least 16 transistors measured for each SWCNT network.

All devices exhibited balanced ambipolar charge transport characteristics (*i.e.*, electron and hole conduction) with low gate leakage currents as shown in the current-voltage (transfer) characteristics (linear regime, source-drain voltage $V_{ds}$ = -0.1 V) in **Figure 3b**. Despite decreasing electron and hole currents with increasing $sp^3$ defect density, even devices with the highest degree of functionalization remained fully functional and showed good switching behavior. Note that the current hysteresis resulted from incomplete removal of oxygen and water, which act as charge traps,[39] due to the relatively low annealing temperatures that were required to avoid defect cleavage.[20]

The increasing number of $sp^3$ defects causes a decrease in charge carrier mobilities for both holes and electrons. **Figure 3c** shows the carrier mobilities normalized to the values for the pristine SWCNT network transistors *versus* the Raman D/G$^+$ area ratio. We find a decrease of ~50–70% for the highest density of $sp^3$ defects compared to the reference transistors. Absolute linear mobilities range between $1.18 \pm 0.05$ cm$^2$ V$^{-1}$ s$^{-1}$ (pristine) to $0.35 \pm 0.01$ cm$^2$ V$^{-1}$ s$^{-1}$ (highest defect density) for electrons, whereas hole mobilities decrease from $4.78 \pm 0.20$ cm$^2$ V$^{-1}$ s$^{-1}$ to $2.26 \pm 0.06$ cm$^2$ V$^{-1}$ s$^{-1}$ upon $sp^3$ functionalization (see also **Supporting Information, Figure S13** and **Table S1**). These values are in excellent agreement with previous results for functionalized SWCNT network FETs with bromoaryl instead of nitroaryl defects.[20]

The electrical measurements of SWCNT FETs (channel length of 20 μm) represent the macroscopic charge transport in nanotube networks involving both intra-nanotube transport and carrier hopping across nanotube-nanotube junctions. As such, the observed decrease in mobilities by ~50–70% for the highest degree of functionalization reflects the interaction of charge carriers with multiple $sp^3$ defects. Note also that FETs operate at high charge carrier densities ($10^{11}$ - $10^{13}$ cm$^{-2}$, corresponding to a few tens to hundreds of charge carriers per μm of SWCNT), and mobilities in SWCNTs increase with carrier concentration before reaching a maximum.[19, 40] Furthermore, the interaction of nanotubes with the substrate (phonon scattering)[41] and with each other affect the overall carrier mobility. All this is in clear contrast to OPTP measurements of individual SWCNTs in dispersion, where a low density of photogenerated carriers is locally driven over a few tens of nm and may scatter at one or only a few defect sites. Taking these differences into account, the ~25 % reduction in carrier mobility (~30 % reduction in photoconductivity) upon *sp*³ functionalization obtained from OPTP measurements indicates a comparable impact of luminescent defects on the microscopic and macroscopic carrier mobilities in functionalized individual SWCNTs and their networks, respectively. Since THz spectroscopy of SWCNT dispersions unambiguously probes only the intrinsic nanotube charge transport, we can assume that the intra-nanotube resistance (increased by the *sp*³ defects) in (6,5) SWCNT networks – in contrast to common belief – is not negligible compared to the junction resistance (inter-nanotube hopping).[42]

To further corroborate the impact of *sp*³ defects on charge transport in individual SWCNTs within networks, we performed temperature-dependent OPTP spectroscopy on dense (6,5) SWCNT films. As shown in **Figure S14a** (**Supporting Information**), for pristine nanotubes at 288 K, the THz photoconductivity of thin films was twice as high as that of SWCNTs in dispersion. This is not surprising, given that the enhanced electronic coupling of nanotubes in networks should facilitate exciton dissociation onto different nanotubes.[43] Inter-nanotube exciton dissociation leads to an increased photon-to-free-charge conversion quantum yield in films compared to dispersions. We further find that the (rescaled) frequency-resolved photoconductivities for film and dispersion show identical spectra (**Figure S14b, Supporting Information**) and thus the same momentum scattering time for a given pump-probe delay. **Figure 4a** shows the photoconductivity dynamics of thin films of pristine and highly *sp*³-functionalized SWCNTs at room temperature. Similar to the dispersions (see **Figure 2**), the peak photoconductivity normalized by the absorbed photon density was ~30 % lower for the functionalized films compared to the pristine reference. The similarity of the photoconductivity change for nanotube dispersions and films indicates that OPTP probes the intra-nanotube transport even in dense networks. We note that the observed trends of photoconductivity changes upon functionalization also persist for higher absorbed photon densities and consequently higher charge carrier densities. These higher carrier densities come close to the low-voltage range in FET measurements (see above) and corroborate the comparability of both techniques.

Measurements at low temperatures (at 78 K in **Figure 4a**, for other temperatures, see **Supporting Information**, **Figure S15**) showed a significant increase of the photoconductivity amplitude for the pristine SWCNT film, whereas $sp^3$-functionalized SWCNTs exhibited an almost negligible increase (**Figure 4b**). We employed temperature-dependent THz-TDS measurements to record complex conductivity spectra (see **Figure 4c** and **Supporting Information, Figure S16**) and analyzed the frequency-resolved spectra by the DS model to extract the carrier mobility as a function of temperature. The observed inverse temperature dependence of the mobility ($\mu \propto 1/T$, see **Figure 4d**) is in agreement with single-nanotube FET measurements and suggests that for pristine SWCNT films charge transport within individual nanotubes is mainly limited by phonon scattering.[15, 19] In contrast to that, charge transport through macroscopic SWCNT networks (*e.g.*, in FETs) is always thermally activated due to carrier hopping across nanotube-nanotube junctions.[40]

As shown in **Figure 4d**, the increase in mobility with decreasing temperatures was more pronounced for pristine SWCNTs than for the $sp^3$-functionalized networks, which is in agreement with the temperature-dependent photoconductivity amplitudes (**Figure 4b**). For both samples, phonon scattering, which is the cause of this temperature dependence,[15, 19] should be reduced to a similar degree at lower temperatures. However, the additional charge scattering at $sp^3$ defects significantly reduces the inverse temperature dependence of the intra-nanotube mobility for functionalized SWCNTs as well as the absolute values.

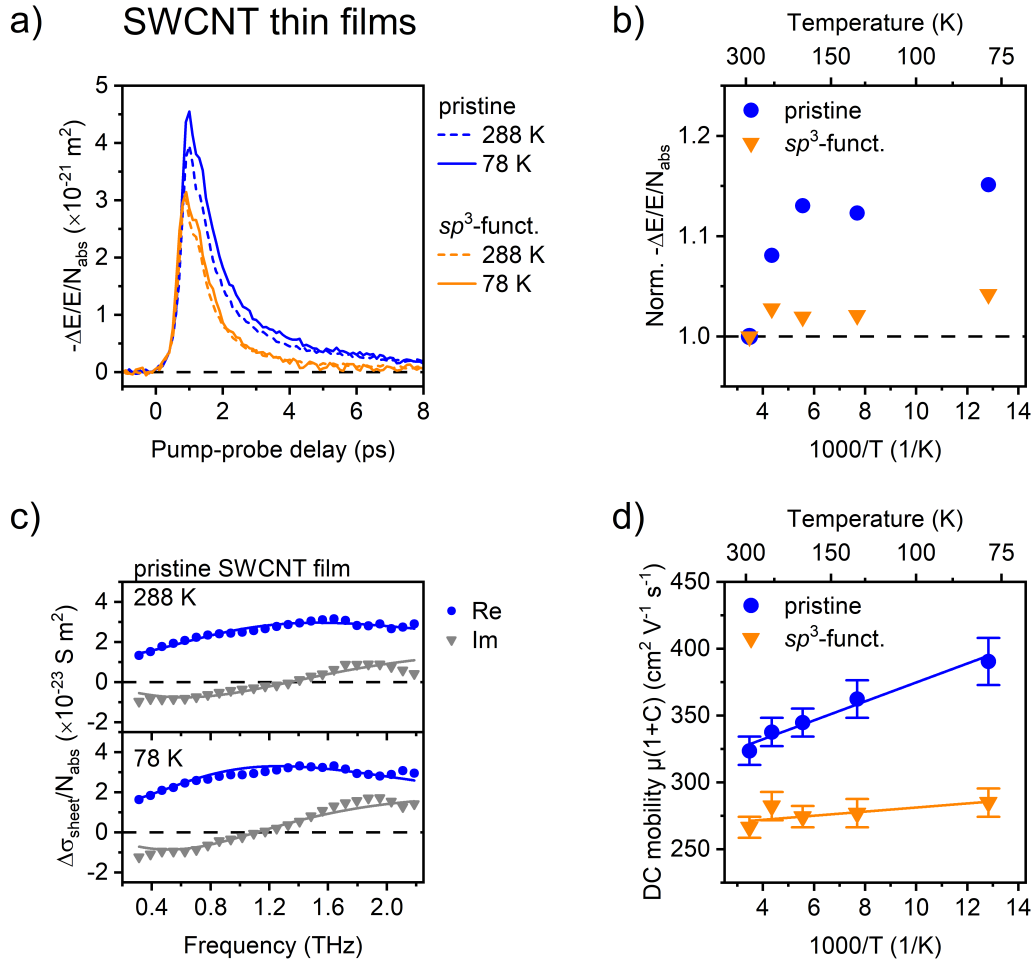

**Figure 4. a)** Temperature-dependent photoconductivity (proportional to –ΔE/E) dynamics of pristine and *sp*[3]-functionalized SWCNT thin films at room temperature (288 K, dashed lines) and 78 K (solid lines). The optical excitation energy was 1.25 eV and the absorbed photon density was ~2·10$^{18}$ m$^{-2}$. **b)** Temperature-dependent maximum photoconductivity normalized to the values at 288 K for pristine (blue circles) and *sp*[3]-functionalized (orange triangles) SWCNT films, respectively. **c)** Complex photoconductivity spectra for the pristine SWCNT thin film at 288 K and 78 K measured ~1.5 ps after photoexcitation (blue circles, real conductivity; gray triangles, imaginary conductivity). The solid lines represent fits to the data with the DS model. **d)** Carrier mobilities in the DC limit extracted from the fits in

c) as a function of inverse temperature (blue circles, pristine SWCNT film; orange triangles, $sp^3$-functionalized SWCNT film). Solid lines are linear fits to the data.

Differences in the temperature dependence of carrier mobilities were observed previously for FETs with pristine and $sp^3$-functionalized (6,5) SWCNT networks: At higher temperatures, the impact of the defects was small, whereas at low temperatures, the presence of $sp^3$ defects significantly increased the temperature dependence of the mobility (*i.e.*, the mobilities decreased much more with decreasing temperature compared to a pristine network).[20] This observation can now be rationalized by the insights into the temperature-dependent intra-nanotube mobilities in $sp^3$-functionalized nanotubes obtained by time-resolved THz spectroscopy. At higher temperatures, thermally activated inter-nanotube carrier hopping dominates the temperature dependence of network mobilities, and the changes in intra-nanotube transport upon $sp^3$ functionalization do not have a significant impact. At low temperatures, however, inter-nanotube hopping is slowed down and the differences in intra-nanotube transport become more apparent. For the pristine SWCNT network, the intra-nanotube mobility increases with decreasing temperature due to reduced phonon scattering (band transport), thus counteracting the decrease in network mobilities due to the reduced carrier hopping. However, for the functionalized SWCNTs, the temperature dependence of intra-nanotube transport is significantly diminished (it barely increases with decreasing temperature, see **Figure 4d**). Therefore, it cannot counteract the decrease in inter-nanotube carrier mobility at lower temperatures, and thus, the overall network mobilities decrease even further at lower temperatures compared to the pristine SWCNT networks. Consequently, the field-effect mobilities of $sp^3$-functionalized SWCNT networks exhibit a stronger temperature dependence than those of pristine (6,5) SWCNT network FETs. Overall, only a convolution of inter-nanotube hopping and intra-nanotube band transport (affected by the $sp^3$ defects) can explain the observed temperature dependence of network mobilities in FETs with functionalized nanotubes. These results provide further evidence that junction resistances are not the only limiting factor for charge transport through random SWCNT networks.

CONCLUSION

We have investigated the impact of luminescent $sp^3$ defects on charge transport in polymer-wrapped (6,5) carbon nanotubes on a microscopic and macroscopic level by combining (temperature-dependent) OPTP and THz time-domain spectroscopy of dispersions and thin films with electrical measurements of network FETs. The complementary nature of these methods allows us to evaluate the contributions of intra-nanotube transport and nanotube-nanotube junctions in SWCNT networks. We find that the introduction of luminescent $sp^3$ defects leads to increased charge scattering along SWCNT segments and reduced carrier mobility, photoconductivity, and lifetime. It also diminishes the intrinsic increase of intra-nanotube carrier mobilities with decreasing temperature. Since the reduction of the field-effect mobilities in functionalized nanotube networks is on the same order of magnitude as that of the photoconductivity, we propose that the apparent carrier mobilities (holes and electrons) in (6,5) SWCNT networks are not only limited by the junctions but result from a superposition of intra- and inter-nanotube contributions.

METHODS

**Selective Dispersion of (6,5) SWCNTs.** (6,5) SWCNTs were selectively extracted from CoMoCAT raw material (Sigma Aldrich, batch no. MKCJ7287, 0.4 g L$^{-1}$) via polymer-wrapping with poly[(9,9-dioctylfluorenyl-2,7-diyl)-*alt*-(6,6'-(2,2'-bipyridine))] (PFO-BPy, American Dye Source, $M_W$ = 40 kg mol$^{-1}$, 0.5 g L$^{-1}$) in toluene under shear force mixing (Silverson L2/Air mixer, 10230 rpm, 72 h) as described previously.[30] To remove residual aggregates and impurities, two consecutive centrifugation steps (60000*g*, 45 min each) and subsequent filtration through a polytetrafluoroethylene (PTFE) syringe filter (pore size 5 µm) were employed. The obtained dispersion was either directly used for *sp*$^3$ functionalization of (6,5) SWCNTs, or filtered through a PTFE membrane (Merck Millipore JVWP, pore size 0.1 µm) and washed with toluene (10 mL) for the fabrication of reference samples as detailed below.

***sp*$^3$ Functionalization of (6,5) SWCNTs.** As described previously,[31] PFO-BPy-wrapped (6,5) SWCNTs were covalently functionalized with 4-nitrobenzenediazonium tetrafluoroborate in a solvent mixture of toluene and acetonitrile (80:20 vol-%). Reactions were performed at a (6,5) SWCNT concentration of 0.54 mg L$^{-1}$, corresponding to an optical density of 0.3 cm$^{-1}$ at the $E_{11}$ absorption peak. An appropriate amount of 18-crown-6 (concentration in the final reaction mixture, 2 mg mL$^{-1}$) was dissolved in toluene and then added to the nanotube dispersion. Subsequently, a solution of the diazonium salt in acetonitrile was added to achieve a concentration in the final reaction mixture between 50 µg mL$^{-1}$ and 750 µg mL$^{-1}$. After thorough mixing, the reaction was allowed to proceed at room temperature in the dark for ~16 h. Then, the reaction mixture was filtered through a PTFE membrane filter (Merck Millipore JVWP, pore size 0.1 µm) and the collected SWCNTs were washed with acetonitrile and toluene in order to remove unreacted diazonium salt as well as excess polymer.

**Preparation of SWCNT Dispersions for FET Fabrication and THz Measurements.** For the fabrication of SWCNT network field-effect transistors (FETs), filter cakes of pristine and $sp^3$-functionalized (6,5) SWCNTs were re-dispersed in 1 mL of fresh toluene *via* bath sonication (30 min) to obtain concentrated dispersions (optical density of 8-10 cm$^{-1}$ at the $E_{11}$ transition) that were immediately used for spin-coating (see below). For THz measurements of SWCNT dispersions, filter cakes were re-dispersed in 1 mL of a PFO-BPy solution in toluene (0.1 g L$^{-1}$) to increase the stability of the dispersions.

**Fabrication of (6,5) SWCNT Network FETs.** On low-sodium glass substrates (Schott AG, AF32eco, 300 µm thickness), interdigitated bottom-contact electrodes ($L$ = 20 µm, $W$ = 10 mm) were patterned by photolithography (LOR5B/S1813 resist, microresist technology GmbH) and electron beam evaporation of chromium (3 nm) and gold (30 nm), followed by lift-off in *N*-methyl pyrrolidone. SWCNT networks were deposited from concentrated dispersions *via* spin-coating (3 × 80 µL, 2000 rpm, 30 s) onto the electrodes with intermediate annealing steps at 120 °C. To remove residual polymer, substrates were subsequently rinsed with tetrahydrofuran and 2-propanol. All SWCNTs outside the channel area were removed by oxygen plasma treatment. After an annealing step at 150 °C for 30 min in dry nitrogen atmosphere, a double-layer dielectric was applied, *i.e.*, ~11 nm of poly(methyl methacrylate) (PMMA, syndiotactic, Polymer Source, $M_\text{W}$ = 315 kg mol$^{-1}$) were spincoated from n-butylacetate followed by atomic layer deposition of ~61 nm of hafnium oxide (Ultratech Inc., Savannah S100) at 100 °C using a tetrakis(dimethylamino)hafnium precursor (Strem Chemicals Inc.) and water as the oxidizing agent. Thermal evaporation of 30 nm silver top-gate electrodes through a shadow mask completed the devices.

**Preparation of (6,5) SWCNT Thin Films for THz measurements.** Filter cakes of pristine and *sp*³-functionalized (6,5) SWCNTs were re-dispersed in 1 mL of fresh toluene *via* bath sonication and diluted to an optical density of ~0.1 cm$^{-1}$ at the $E_{11}$ absorption transition. Of each dispersion, 10 mL were filtered over respective mixed cellulose ester membranes (Merck Millipore VSWP, pore size 0.025 µm). Cleaned fused silicon dioxide substrates (from PI-KEM Ltd.) were subjected to UV/ozone treatment. Filter membranes with SWCNT films were cut to the desired size and placed onto the substrates with the SWCNT film facing down. The filter cake was wetted with 2-propanol, pressed against the substrate, and the sample was immediately put into an acetone bath (7 × 15 min) to dissolve the filter membrane to leave only the SWCNT film on the substrate. Samples were rinsed with acetone and 2-propanol and then blow-dried with nitrogen.

**Characterization.** Baseline-corrected absorption spectra of SWCNT dispersions were acquired with a Cary 6000i UV-vis-nIR spectrometer (Varian Inc.). Raman spectra (excitation wavelength 532 nm) were measured on drop-cast SWCNT films using a Renishaw inVia confocal Raman microscope with a 50× long working distance objective (Olympus, N.A. 0.5). PL spectra were measured on a home-built laser setup under $E_{22}$ excitation (575 nm) with the wavelength-filtered output of a picosecond-pulsed supercontinuum laser (Fianium WhiteLase SC400, 20 MHz repetition rate). The laser was focused onto the samples with a nIR-optimized 50× objective (Olympus, N.A. 0.65). Scattered laser light was blocked by appropriate longpass filters, and emission spectra were recorded with a grating spectrometer (Acton SpectraPro SP2358, 150 lines mm$^{-1}$) equipped with a liquid nitrogen-cooled InGaAs line camera (Princeton Instruments OMA V:1024-1.7 LN). Atomic force micrographs were acquired with a Bruker Dimension Icon under ambient conditions. Current-voltage characteristics of SWCNT network

FETs were measured in inert atmosphere with an Agilent 4156C semiconductor parameter analyzer.

**THz Spectroscopy.** THz spectroscopy was performed with a commercial, regenerative amplified, mode-locked Ti:sapphire femtosecond laser system from Spectra Physics Spitfire Ace. The laser system provides characteristic 1 kHz pulses of approximately 50 fs duration and a central wavelength of 800 nm. THz waves were generated by optical rectification on a zinc telluride (ZnTe) crystal (along the ⟨110⟩ orientation). The transmitted THz wave was sampled by a second 800 nm pulse *via* the electro-optic effect by a second ZnTe detection crystal. To track the dynamics of the real photoconductivity in optical-pump THz-probe (OPTP) measurements, the pump-induced THz absorption was measured by fixing the sampling beam to the peak of the THz field. By varying the time delay between the pump and sampling beam with an optical delay stage, the time-dependent photoconductivity was recorded. The pump path contained an optical parametric amplifier and mixing stages (Light Conversion, TOPAS) to convert the incident 800 nm light to ~1000 nm wavelength light. During the measurements, the entire THz set-up was kept under dry nitrogen atmosphere to avoid THz absorption. Samples were either purged with dry nitrogen or placed under vacuum conditions ($< 2 \cdot 10^{-4}$ mbar).

AUTHOR INFORMATION

**Corresponding Authors**

*E-mail: wanghai@mpip-mainz.mpg.de

*E-mail: zaumseil@uni-heidelberg.de

**Author Contributions**

# W.Z. and N.F.Z. contributed equally.

ACKNOWLEDGMENTS

This project has received funding from Max Planck Society and the European Research Council (ERC) under the European Union's Horizon 2020 research and innovation programme (Grant agreement no. 817494, "TRIFECTs").

# Supporting Information

**Contents**

# Details for Optical-Pump THz-Probe and THz Time-Domain Spectroscopy

**Extraction of Frequency-Resolved Photoconductivity**

To gain insight into the nature and transport properties of charge carriers, we conducted THz time-domain spectroscopic (THz-TDS) measurements at a fixed pump-probe delay time. We sampled the entire THz waveform transmitted through the sample without photoexcitation as reference (*E*), and photo-induced THz waveform modulation Δ*E* = (*E*<sub>pump</sub> - *E*) in the time domain. Wait — following the guidelines: photo-induced THz waveform modulation $\Delta E = (E_{pump} - E)$ in the time domain. For the thin-film geometry, we converted the relative change of the THz transmission from time into frequency domain following Fourier transformation, and further quantified the complex sheet photoconductivity spectra according to the thin film approximation:[1,2]

$$\Delta\sigma_{\text{sheet}}(\omega) = -\frac{\Delta E(\omega)}{E(\omega)} \cdot \frac{n_1+n_2}{Z_0} \qquad (1)$$

where $Z_0 = 377\ \Omega$ is the impedance of free space, $n_1$ and $n_2$ are the refractive indices of the media before and after the sample, respectively.

Now we consider the frequency-resolved photoconductivity of SWCNTs dispersion in organic solvents (see data in **Figure 2d**). From our measurement geometry, the THz probe field travels from air into the front cuvette window, the solvent, and then the back of the cuvette window and back into air. Due to the low concentration of the SWCNT dispersion, the penetration depth of the excitation is larger than the dispersion thickness. The excitation profile through the entire suspension is therefore treated to be homogeneous. The complex photoconductivity $\Delta\sigma(\omega)$ follows:[2,3]

$$\Delta\sigma(\omega) = \left(n^2 - (1-\Delta\hat{n}(\omega))^2\right)\frac{i\omega}{Z_0 c}, \quad \text{with}\ \Delta\hat{n}(\omega) = \left(\frac{1}{n}\frac{n_w-n}{n_w+n} + i\frac{\omega d}{c}\right)^{-1}\frac{\Delta E(\omega)}{E} \qquad (2)$$

where $n$ and $n_\omega$ are the real refractive indices of the solvent and cuvette, respectively. $\omega$, $d$, and $c$ denote the angular frequency, excitation thickness, and the speed of light. Knowing the refractive indices of the cuvette's windows and the solvent, one can extract the complex photoconductivity in a dispersion from the measured reference THz field and the photoinduced modulation.

# Characterization of (6,5) SWCNT Dispersions for OPTP Spectroscopy

**Absorption Spectra**

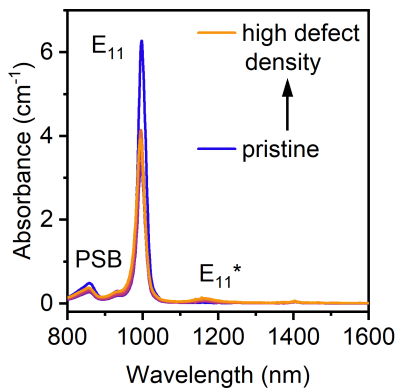

**Figure S1.** Absolute absorption spectra of concentrated pristine and $sp^3$-functionalized (6,5) SWCNT dispersions for optical-pump THz-probe measurements.

**$sp^3$ Defect Density Metrics**

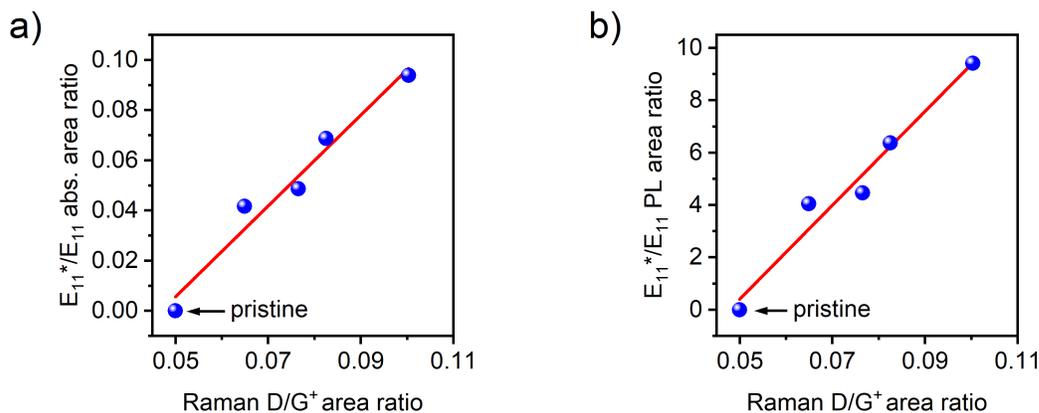

**Figure S2.** Defect density metrics of pristine and $sp^3$-functionalized (6,5) SWCNT dispersions for optical-pump THz-probe measurements. **a)** $E_{11}^*/E_{11}$ absorption peak area ratio and **b)** $E_{11}^*/E_{11}$ PL peak area ratio versus Raman $D/G^+$ area ratio for pristine (6,5) SWCNT dispersions and dispersions of $sp^3$-functionalized (6,5) SWCNTs with different defect densities. Data points for the pristine samples are indicated. Red solid lines are linear fits to the data.

**AFM Length Statistics of Different SWCNT Dispersions** <span style="float:right">S-5</span>

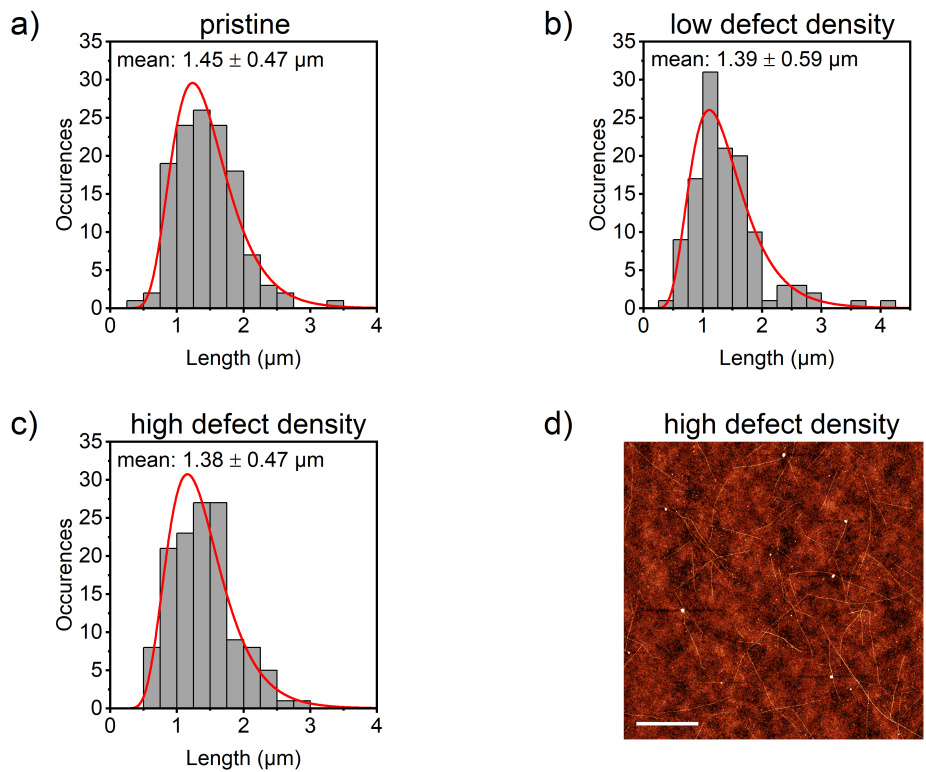

**Figure S3. a)-c)** AFM statistics on individual SWCNT lengths (>120 nanotubes counted each) show that the average nanotube length is not significantly altered upon functionalization. Red lines are log-normal distributions. **d)** Atomic force micrograph (5 × 5 µm) of individual nanotubes with a high $sp^3$ defect density. Scale bar is 1 µm.

# Additional Data for OPTP Spectroscopy of SWCNT Dispersions

**Complete Datasets of Photoconductivity Dynamics**

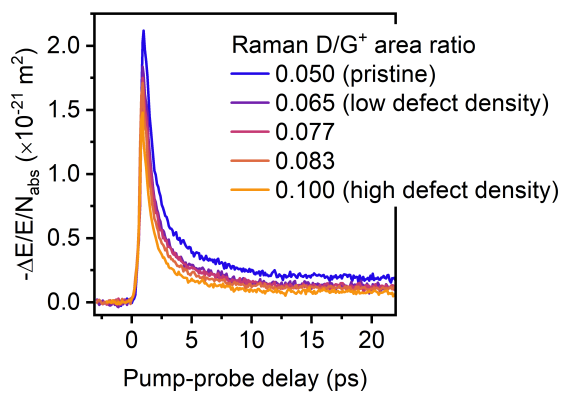

**Figure S4.** Photoconductivity dynamics of pristine and different $sp^3$-functionalized SWCNT dispersions in toluene following optical excitation at 1.25 eV at room temperature.

**Concentration-Dependent Photoconductivity Dynamics**

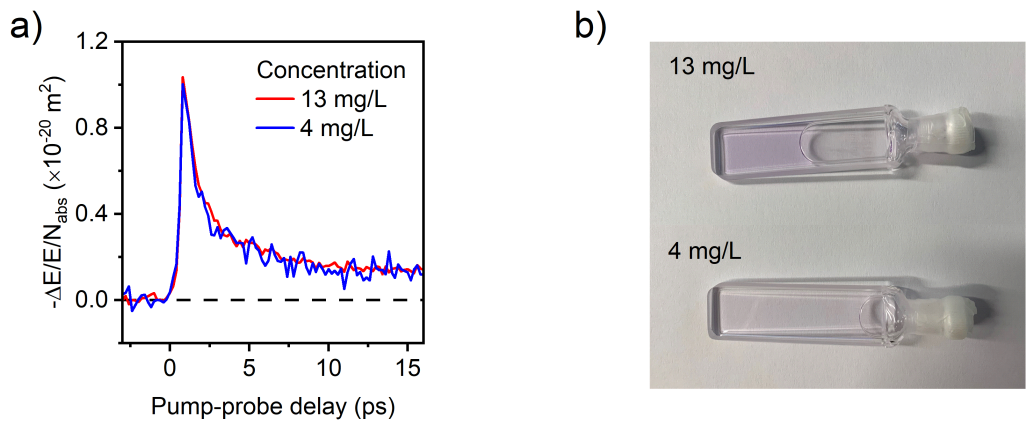

**Figure S5. a)** Photoconductivity dynamics of pristine (6,5) SWCNT dispersions with concentrations of ~13 mg·L$^{-1}$ and ~4 mg·L$^{-1}$. **b)** Photographs of the respective dispersions in cuvettes with 1 mm path lengths. The purple color arises from the $E_{22}$ absorption of (6,5) SWCNTs at 575 nm.

**Carrier Diffusion Length**

The charge carrier diffusion length ($L_D$) was estimated as follows:[4, 5]

$$L_D = \sqrt{\frac{\mu \cdot k_B \cdot T \cdot \tau}{e}} \qquad (3)$$

where $\mu$, $k_B$, $e$, and $T$ are carrier mobility, Boltzmann constant, elementary charge, and temperature, respectively. The carrier lifetime $\tau$ was inferred by fitting to the photoconductivity dynamics (see **Figure S4**).

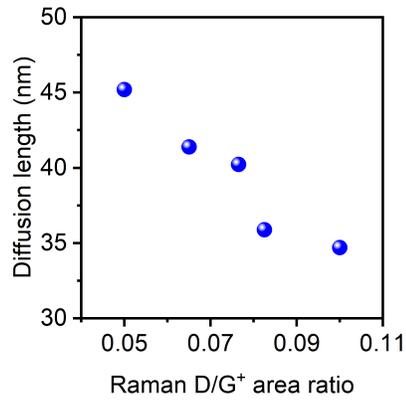

**Figure S6.** Estimated charge carrier diffusion lengths as a function of the *sp³* defect density (Raman D/G⁺ area ratio).

## Normalized Photoconductivity Dynamics and Photoconductivity Lifetimes

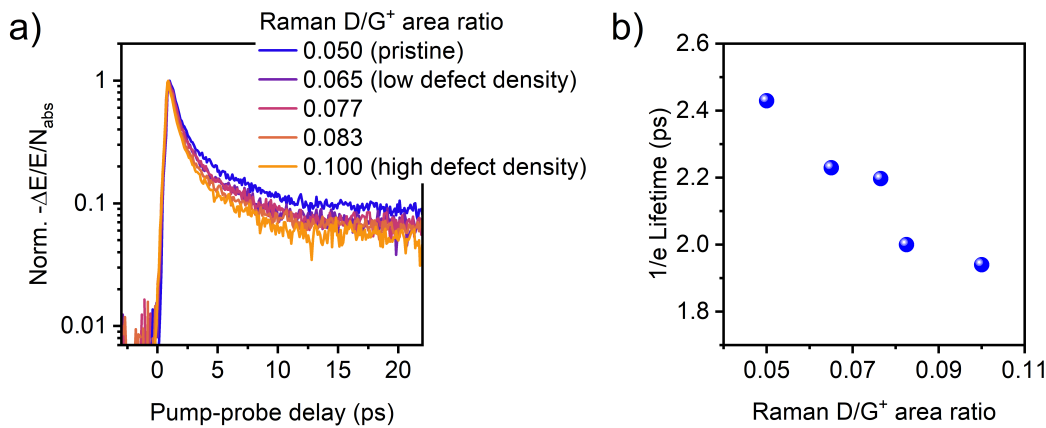

**Figure S7. a)** Normalized photoconductivity dynamics of pristine and *sp*$^3$-functionalized SWCNT dispersions and **b)** photoconductivity 1/e lifetimes, extracted from the data in a).

## Frequency-Resolved Complex Photoconductivity Spectra at Different Time Delays

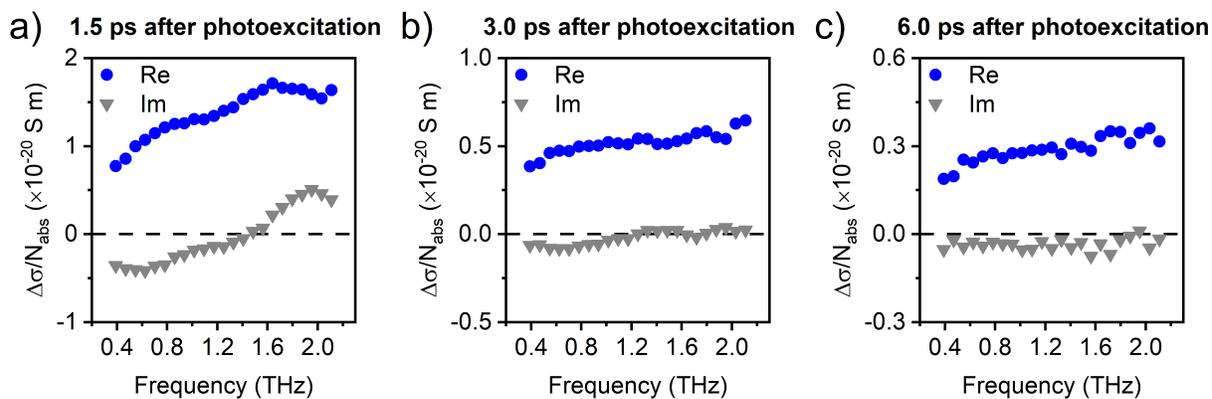

**Figure S8.** Frequency-resolved, complex photoconductivity spectra of pristine (6,5) SWCNT dispersions measured at different time delays (**a-c**) after photoexcitation (blue circles, real conductivity; gray triangles, imaginary conductivity).

**Charge Carrier Scattering Rate**

In **Figure S9a** the charge carrier scattering rate $\gamma$ (that is, the inverse of the effective scattering time $\tau_{\text{eff}}$, $\gamma = \frac{1}{\tau_{\text{eff}}}$) is shown as a function of the $sp^3$ defect density (represented by the Raman D/G$^+$ area ratio). Here, $\tau_{\text{eff}}$ is defined by considering the carrier mobility in the DC limit defined as:

$$\mu_{dc} = \frac{e\tau_{\text{eff}}}{m^*} = \frac{e\tau}{m^*}(1+C) \text{ with } \tau_{\text{eff}} = \tau(1+C) \tag{4}$$

where $m^*$ is the effective mass, $\tau$ is the scattering time, and $C$ is the backscattering rate from the Drude-Smith model fitting.

To further analyse the $sp^3$ defect density contribution to the overall charge scattering frequency, we consider charge scattering contributions from both phonons and defects following Matthiessen's rule at room temperature:[4]

$$\gamma_{\text{total}}(n_d) = \gamma_{\text{phonon}} + \gamma_{\text{defect}}(n_d) \tag{5}$$

or

$$\gamma_{\text{defect}}(n_d) = \gamma_{\text{total}}(n_d) - \gamma_{\text{phonon}} \tag{6}$$

where $\gamma_{\text{total}}$ is the total scattering rate (which is assumed to be equal to $\gamma_{\text{eff}}$), and $\gamma_{\text{phonon}}$ and $\gamma_{\text{defect}}$ are the scattering rates due to phonons and defects, respectively. $n_d$ represents the total defect density in SWCNTs (reflected by the Raman D/G$^+$ area ratio). By assuming that $\gamma_{\text{phonon}}$ is independent of the defect density, we can infer the additional scattering rate $\Delta\gamma$ due to the increase of the $sp^3$-defects as shown in **Figure S9b**. The pristine SWCNTs are used as the reference, and we define $\Delta\gamma = \gamma_{\text{defect}}(n_d) - \gamma_{\text{pristine}}$.

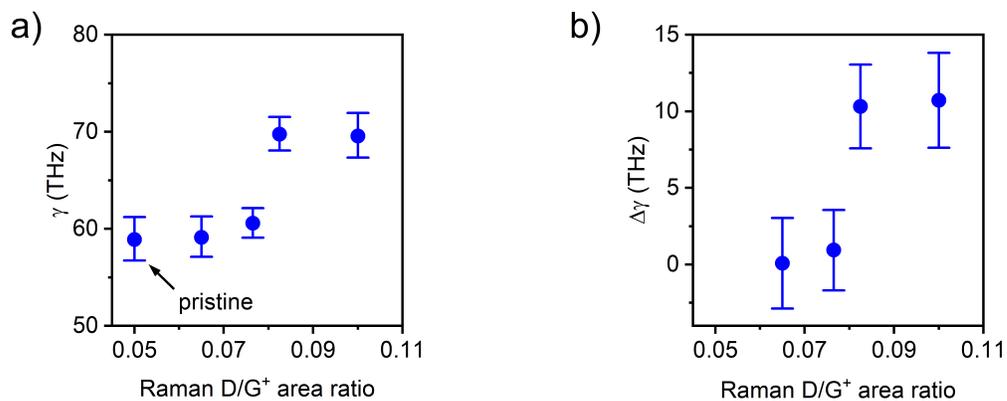

**Figure S9. a)** Charge carrier scattering rate $\gamma$ determined from THz measurements of (6,5) SWCNT dispersions as a function of $sp^3$ defect density (*i.e.*, Raman D/G$^+$ area ratio) and **b)** estimated differences in scattering rates ($\Delta\gamma$) due to the introduction of different concentrations of $sp^3$ defects.

# Additional Data for SWCNT Network FETs

**Atomic Force Micrographs of SWCNT Networks**

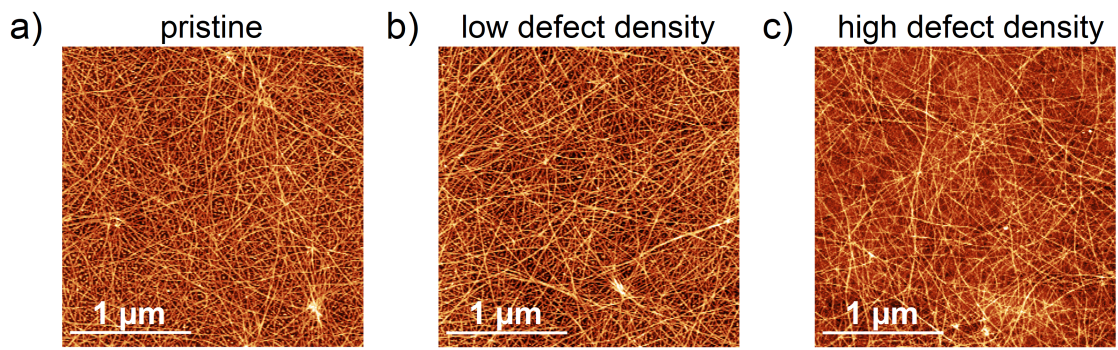

**Figure S10. a)-c)** Atomic force micrographs of the channel areas of SWCNT network FETs. Scale bars are 1 µm.

**Comparison of Samples for OPTP Measurements and FET Fabrication**

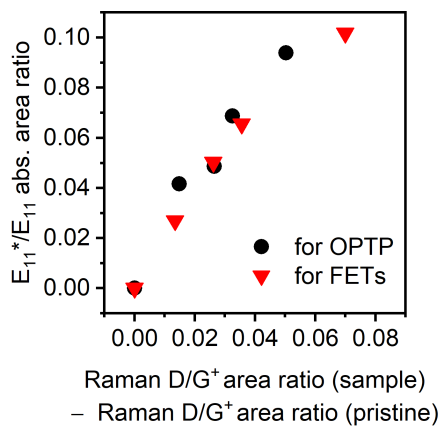

**Figure S11.** Characterization of pristine and functionalized (6,5) SWCNT dispersions for optical-pump THz-probe (OPTP) measurements (black circles) and for FET fabrication (red triangles). The $E_{11}^*/E_{11}$ absorption peak area ratios versus Raman D/G$^+$ area ratios are very similar for both sample series, corroborating a similar density of $sp^3$ defects. Note that on the x-axis, the differential Raman D/G$^+$ area ratio is plotted in order to account for variations in nanotube quality (*i.e.*, number of processing defects) of the starting dispersions for $sp^3$ functionalization and thus different initial Raman D/G$^+$ ratios.

**Spectroscopic Characterization of SWCNT Dispersions for FET Fabrication**

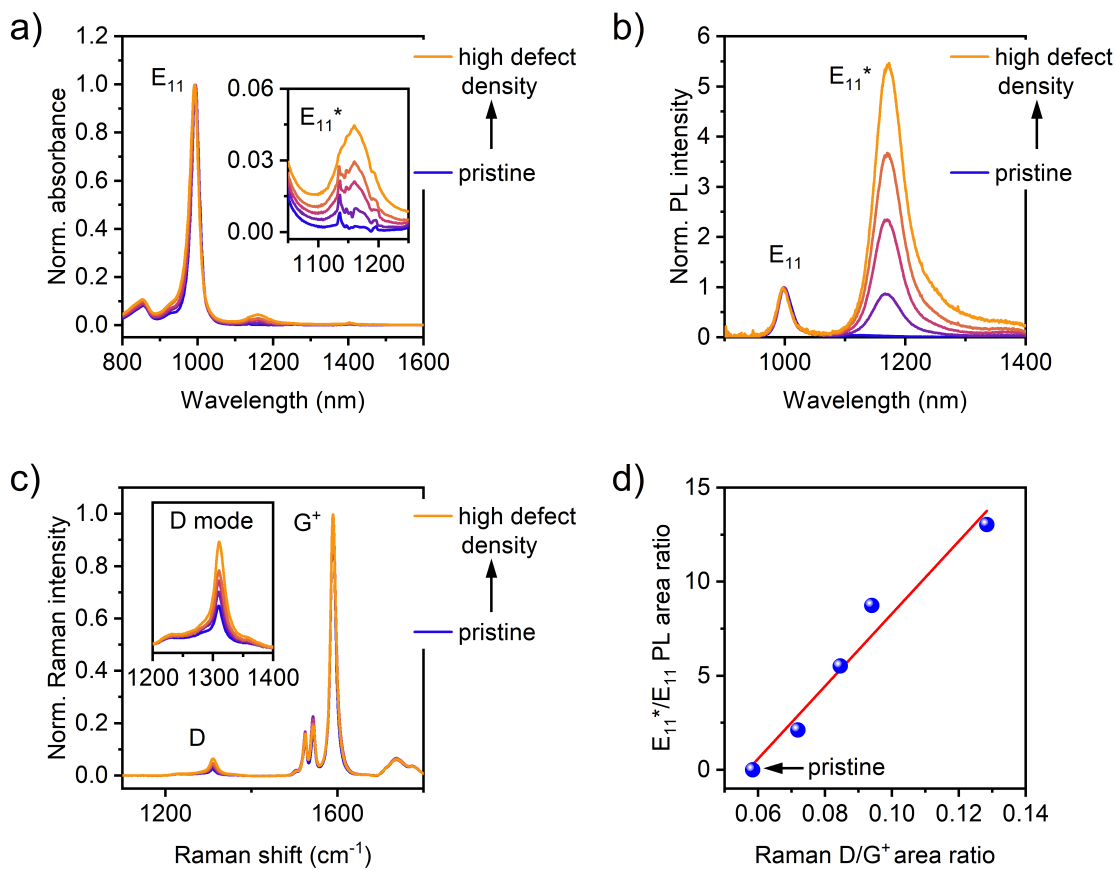

**Figure S12.** Characterization of pristine and functionalized (6,5) SWCNT dispersions for FET fabrication. **a)** Absorption spectra normalized to the $E_{11}$ transition and zoom-in on the $E_{11}^*$ defect state absorption band (inset). **b)** PL spectra ($E_{22}$ excitation at 575 nm) normalized to the $E_{11}$ peak. **c)** Raman spectra normalized to the $G^+$ mode. The inset shows the defect-related D mode that increases with defect density. **d)** $E_{11}^*/E_{11}$ PL area ratio versus Raman $D/G^+$ area ratio for pristine and $sp^3$-functionalized (6,5) SWCNT dispersions with different defect densities. The data point for the pristine sample is indicated. The red solid line is a linear fit to the data.

**Linear Field-Effect Mobilities**

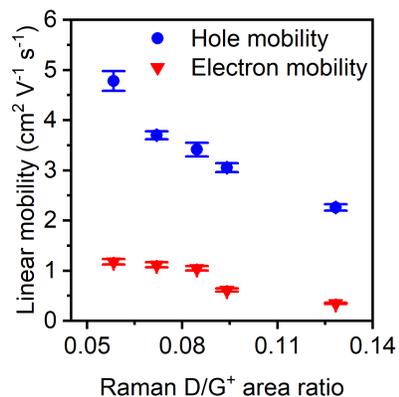

**Figure S13.** Absolute linear field-effect mobilities (holes, blue circles; electrons, red triangles) of pristine and *sp*$^3$-functionalized (6,5) SWCNT networks versus Raman D/G$^+$ area ratio. Error bars are standard deviations for at least 16 transistors measured for each SWCNT network.

# Additional Data for OPTP Spectroscopy of SWCNT Thin Films

**Comparison of the THz Response of SWCNT Dispersions and Thin Films**

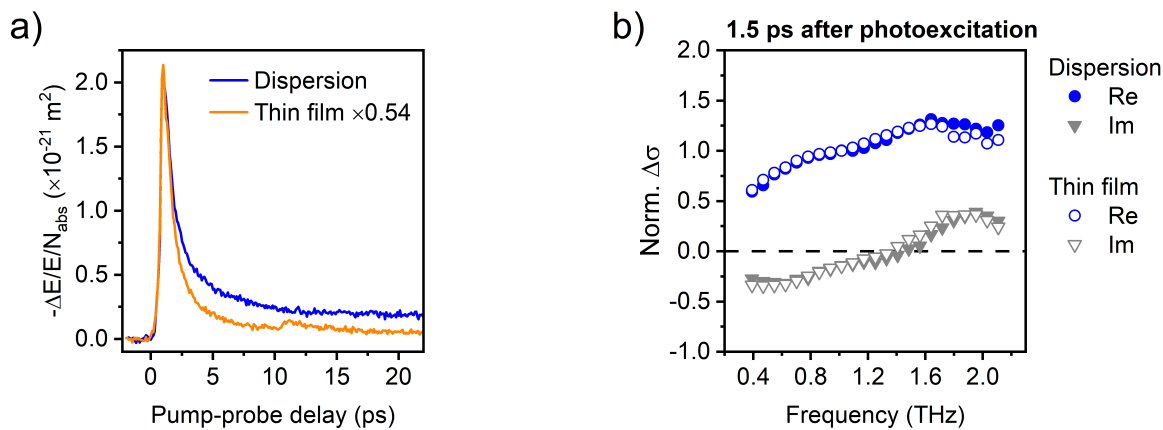

**Figure S14. a)** Photoconductivity dynamics of a pristine (6,5) SWCNT dispersion and thin film following optical excitation at 1.25 eV at room temperature. **b)** Normalized complex photoconductivity spectra of a pristine (6,5) SWCNT dispersion (filled symbols) and thin film (empty symbols) measured ~1.5 ps after photoexcitation (blue circles, real conductivity; gray triangles, imaginary conductivity). The data are normalized by the values at 1 THz.

**Temperature-Dependent Photoconductivity Dynamics**

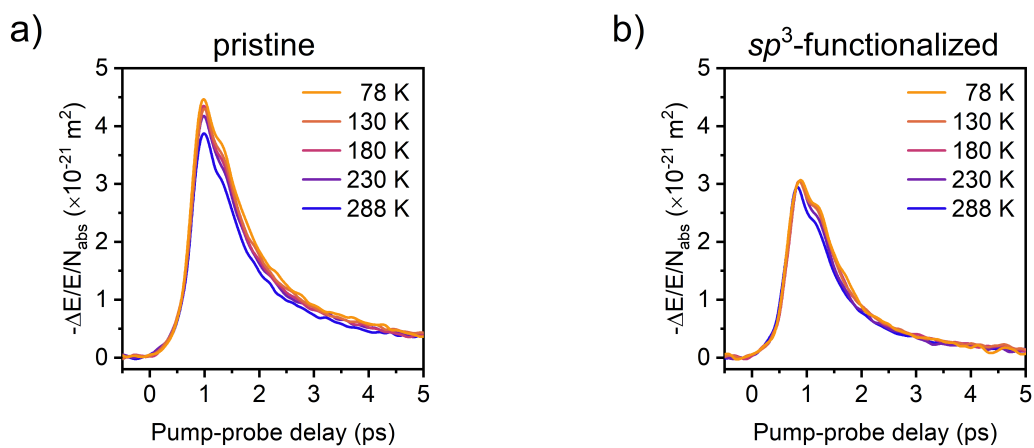

**Figure S15.** Temperature-dependent photoconductivity (which is proportional to $-\Delta E/E$) dynamics of **a)** pristine and **b)** $sp^3$-functionalized SWCNT thin films.

**Temperature-Dependent Complex Photoconductivity Spectra** 

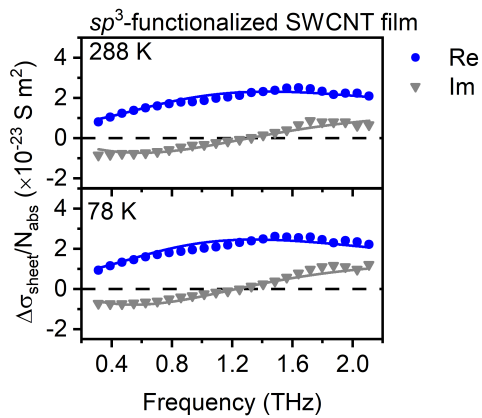

**Figure S16.** Complex photoconductivity spectra for the *sp*³-functionalized SWCNT thin film at 288 K and 78 K measured at ~1.5 ps after photoexcitation (blue circles, real conductivity; gray triangles, imaginary conductivity). The solid lines represent fits to the data according to the Drude-Smith model.

## Overview of Extracted Data from OPTP Spectroscopy and FET Measurements

**Table S1.** Raman D/G$^+$ area ratios, THz photoconductivities, carrier mobilities and lifetimes, and FET mobilities for all samples.

| Sample | OPTP Spectroscopy of SWCNT Dispersions | | | | SWCNT Network FET Characterization | | |
|---|---|---|---|---|---|---|---|
| | Raman D/G$^+$ area ratio | Max. $-\Delta E/E/N_{abs}$ ($\times 10^{-21}$ m$^2$) | Carrier mobility (cm$^2$ V$^{-1}$ s$^{-1}$) | Photo-conductivity 1/e lifetime (ps) | Raman D/G$^+$ area ratio | Hole mobility (cm$^2$ V$^{-1}$ s$^{-1}$) | Electron mobility (cm$^2$ V$^{-1}$ s$^{-1}$) |
| Pristine | 0.050 | 2.12 | 325 ± 8 | 2.43 | 0.058 | 4.78 ± 0.20 | 1.18 ± 0.05 |
| Low defect density | 0.065 | 1.84 | 297 ± 7 | 2.23 | 0.072 | 3.70 ± 0.08 | 1.12 ± 0.05 |
| Medium-low defect density | 0.077 | 1.75 | 285 ± 7 | 2.20 | 0.085 | 3.41 ± 0.14 | 1.05 ± 0.04 |
| Medium-high defect density | 0.083 | 1.71 | 249 ± 6 | 2.00 | 0.094 | 3.05 ± 0.09 | 0.61 ± 0.03 |
| High defect density | 0.100 | 1.48 | 240 ± 5 | 1.94 | 0.128 | 2.26 ± 0.06 | 0.35 ± 0.01 |